# Atomic-scale modeling of the thermal decomposition of titanium(IV)-isopropoxide

Benazir Fazlioglu Yalcin[1,*], Dundar E. Yilmaz[2], Adri CT van Duin[2], Roman Engel-Herbert[1,3]

[1] *Department of Materials Science and Engineering, Pennsylvania State University, University Park, PA, USA*

[2] *Department of Mechanical Engineering, Pennsylvania State University, University Park, PA, USA*

[3] *Paul-Drude-Institute for Solid-State Electronics, Hausvogteiplatz 5-7, 10117 Berlin, Germany*

## ABSTRACT

The metal-organic (MO) compound titanium(IV)-isopropoxide (Ti(O$^i$Pr)$_4$, TTIP) has tremendous technological relevance for thin film growth and coating technologies, offering a low-temperature deposition route for titania and titanium-oxide-based compounds. Thermal decomposition via the release of organic ligands – a key process in any TTIP-based synthesis approach – is commonly assumed to take place only via the *β*-hydride elimination process. Here, we present reactive force field molecular dynamics (ReaxFF-MD) and metadynamics simulations that challenge this conventionally assumed scenario by revealing different, energetically preferred reaction pathways. The initial organic ligand pyrolysis step was spontaneous and occurred exclusively by dissociating a C-O bond, but not always via the *β*-hydride elimination process which suggests other reaction pathways also play a significant role in the decomposition. Among all the following events, C-O bond dissociations occurring without *β*-hydride elimination and Ti-O bond dissociations required hydrogenation of the remaining Ti-containing MO fragment. The complete reaction scheme for the TTIP thermolysis, along with the statistics for the different ligand liberation steps and the associated reaction barriers for the bond dissociation events is presented. ReaxFF-MD simulations performed in the dilute limit realistically capture typical thin film deposition conditions, which in combination with metadynamics data, which produces free energies, constitutes a very powerful tool to quantitatively analyze the reaction dynamics of MO-based thin film growth processes and provide an atomic-scale understanding of how the remaining organic ligands detach from different titanium-containing MO fragments. The approach presented here allows for effective and straightforward identification of the undesirable temperature biasing effects in ReaxFF-MD and represents a predictive framework to identify chemical reaction pathways relevant to film growth processes at the atomic scale under realistic, experimentally relevant conditions. It enables computationally informed engineering of MO molecules with tailored decomposition and reaction pathways, and thus rapid and cost-effective advancements in MO molecule design for existing and future applications of thin film deposition and coating processes.

*To whom correspondence should be addressed. E-mail: bmf5511@psu.edu

## 1. Introduction

Metal alkoxides have been widely used as metal-organic (MO) precursor compounds for synthesizing metal oxides, glass, ceramic compounds, and thin films using different deposition techniques.[1–3] For example, the high-k metal gate process, a much-needed breakthrough that has enabled continued scaling of the Si-based complementary metal-oxide-semiconductor (CMOS) technology, requires the growth of few nm-thick amorphous hafnia layers utilizing hafnium-containing MO precursor molecules and has opened the door for continued scaling of transistor technology.[4] Furthermore, titanium-containing oxide materials, such as high-quality $Pb(Mg_{1/3}Nb_{2/3})O_3$-$PbTiO_3$ (PMN-PT) thin films, that have been integrated epitaxially on vicinal (001) Si revealed giant piezoelectric responses, allow us to fabricate microelectromechanical systems (MEMS)[5] that are being utilized in ultrasound medical imaging, microfluidic control, mechanical sensing, and energy harvesting applications[5,6]. Moreover, $TiO_2$ thin films have tremendous industrial applications in optoelectronics, photovoltaics, as well as pigment industry, and waste treatment.[7–10] This has motivated a large-scale use of the MO precursor titanium(IV) isopropoxide ($Ti(O^iPr)_4$ – or TTIP[11] – as one of the most extensively used MO substances for similar purposes.[12]

TTIP has furthermore come into focus as an indispensable component of a recently developed thin film growth approach called hybrid molecular beam epitaxy (*h*MBE).[13-14] Here, elements evaporated from effusion cells and MOs supplied from a gas injector are co-supplied, combining the advantages of physical and chemical vapor deposition techniques and enabling a self-regulated growth regime[15–19]. This has been proven to provide a pathway towards the highest quality oxide-based materials,[20,21] and allows to scale film growth rates while maintaining stoichiometric control[22], constituting a critical step toward developing semiconductor-grade complex oxide thin films.

Metal alkoxides have the general chemical formula, $[M(OR)_x]_n$, where M represents a metal-cation with valency x, and R is an alkyl group attached to the metal via an oxygen atom.[23] The MO compound has a much higher vapor pressure compared to the elemental metal, allowing it to bring them into their gas phase at much lower temperatures and simplifying their supply during the film deposition process. Specifically, while Ti metal has a vapor pressure of about 7.5 mTorr close to its melting point at 1660 °C[24], TTIP has a vapor pressure of 100 mTorr already at room temperature.[13] Sufficient film deposition temperatures are required to allow the thermal decomposition of the MO into the non-volatile fragment containing the element of interest and volatile fragments that are being pumped away and do not get incorporated into the film. Furthermore, the weak bonding of the TTIP molecule with the film surface allows long diffusion lengths of TTIP compared to the very reactive elemental Ti, allowing it to access desired layer-by-layer growth modes at much lower temperatures.[13]

In the case of the TTIP molecule, a single Ti is tetrahedrally coordinated by oxygen as part of the isopropoxy group surrounding the central Ti atom, see Fig. 1. The onset of thermal decomposition of TTIP has been reported to occur around 250°C making TTIP deposition process temperature compatible with back-end-of-line processing.[25–27] When subject to sufficient energy, e.g. at elevated temperatures,[13,25–27] TTIP has been experimentally observed to decompose into $TiO_2$, highly volatile organic by-products, and water: $Ti(OC_3H_7)_4 \rightarrow TiO_2 + 4C_3H_6 + 2H_2O$.[27] The underlying reaction mechanism is assumed to be the *β*-hydride elimination process, in which an alkyl group bonded to the metal center atom is dissociated and converted into the corresponding alkene[28] that has been claimed to be a vital step in the thermal dissociation of MO substances.[29–31]

Specifically, for a decomposing TTIP molecule one of the propyl groups bonded to oxygen is assumed to lose a H atom from a *β*-carbon to form propene while leaving a three-ligand TTIP with one less ligand group behind. In the meantime, the released H usually re-attaches to the remaining TTIP fragment molecule by forming a bond with the oxygen that the corresponding alkyl group was originally bonded to.

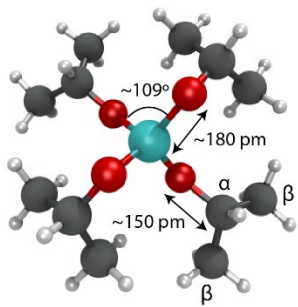

Figure 1. A single TTIP molecule in a geometrical configuration after energy minimization based on ReaxFF using pseudopotentials of Ti, O, C, and H.[32–35] The center atom (blue) is titanium, which is coordinated tetrahedrally by oxygen atoms (red) tying the organic ligands comprised of carbon (black) and hydrogen (white) atoms to the center metal cation. Typical bond lengths of the C-O and the Ti-O bonds are given along with the angle between Ti-O bonds showing a tetrahedral arrangement of the alkyl groups. The middle C atom through which the C-O bond is established is referred to as *α*-carbon and the neighboring C atom is referred to as *β*-carbon.

The utilization of MO compounds for the synthesis of thin films and coatings is generally well established and widely employed in the lab for the development of new thin film synthesis recipes[13,14,36,37] as well as on an industrial scale to produce titania,[38] hafnia,[39] and gallium-containing compound semiconductors, such as GaAs.[40] However, only little is known about the reaction kinetics of the decomposition processes, in particular about the by-product formation and their role in aiding or hindering the overall reaction. Only a few studies have been carried out to elucidate the decomposition pathways of TTIP using a quantum mechanical description via density functional theory (DFT) calculations.[25,41,42] Specifically, Buerger *et al.* calculated the optimized ground state geometries of Ti-containing species, predicted the thermodynamic properties for TTIP and some selected decomposition products and reported the numeric results of the formation enthalpy, entropy and heat capacity.[41] In a follow-up study, the analysis was expanded towards the kinetics for the thermal decomposition of TTIP, in which the main intermediate species of TTIP that were produced under combustion conditions and hydrolysis reactions were analyzed.[25] Another study focused on the UV-photodissociation of the TTIP molecules and reported that under such high energy conditions, the organic ligands were liberated via an acetone elimination reaction as the primary TTIP decomposition process.[42] As a ReaxFF-MD study, Wei *et al.* investigated titania nanocluster formation and simulated the TTIP hydrolysis reaction in the high-pressure limit as the primer stage by assuming the presence of water as an oxidizing agent and a large TTIP molecule density in the gas phase that exceeded atmospheric pressures.[43]

The gas phase reaction of TTIP molecules in the dilute limit, a realistic representation of experimental conditions in chemical vapor deposition processes, has not yet been studied. Here, we report on TTIP decomposition analysis using the ReaxFF molecular dynamics (ReaxFF-MD) approach.[43–45] First, the decomposition of 25 TTIP molecules has been analyzed using several simulations under similar initial conditions to improve the statistical significance of the results. After realizing that previously released fragments from ligands can affect the overall reaction process, the decomposition of isolated TTIP molecules was analyzed by exposing them to typical reactants formed during the decomposition process. Specifically, H, $H_2$, and $H_2O$ were added to clarify the effect of these by-products on the overall TTIP decomposition kinetics. Although *β*-hydride elimination was observed in some of the decomposition steps, as it has been assumed to be the dominant reaction process so far,[43,46] it was found that it was never the only and not always the dominant mechanism for the TTIP decomposition. In addition, metadynamics simulations were performed to quantify the free energy evolution of the different chemical reactions that occurred during ReaxFF-MD simulations. The Ti-O bond energies were found to be higher compared to C-O bonds irrespective of TTIPs fragmentation stage, which potentially helps to avoid unintentional incorporation of carbon atoms into the growing films. This combined approach of ReaxFF-MD and metadynamics simulations outlined here allows for identifying reaction pathways relevant to thin film growth and coating processes and can be easily expanded towards a much wider selection of MO precursor and growth environments, including the growth front of the forming film. It enables a computationally informed engineering strategy to design MO molecules and can ultimately help in accelerating the development and deployment of new MO precursor molecules for their use in all chemical vapor deposition processes.

## 2. Results and Discussion

TTIP molecules were simulated in ReaxFF using a previously developed and refined parameter set for Ti, C, O, H.[32–35] A typical TTIP molecule is shown in Fig.1 with a Ti atom at its center that is tetrahedrally coordinated by four isopropoxy groups. Each MO ligand is linked to the titanium atom via an oxygen atom attached to an α-carbon. The Ti-O and C-O bond lengths are about 180 pm and 150 pm, respectively. Initially, 25 TTIP molecules were simulated in a sufficiently large volume to correctly represent the dilute limit inherent to chemical vapor phase deposition processes and to get an overview of the entire thermal decomposition process to discuss the main reaction steps in detail. Next, the decomposition pathway in the presence of particular by-products was studied using a single, isolated TTIP molecule. Since H atoms were found to play a critical role in the thermal decomposition process, we also performed ReaxFF simulations by adding 10 H, 5 $H_2$, and 5 $H_2O$ molecules to a TTIP molecule and compared them with the thermal decomposition of a single TTIP molecule in the absence of any additional molecules. Finally, metadynamics simulations have been utilized to analyze the activation barrier for the individual bond-breaking events which were compared to the statistics previously obtained by ReaxFF-MD simulations. An energy-minimized TTIP molecule was used for all simulation types.

Step-wise thermal decomposition of TTIP

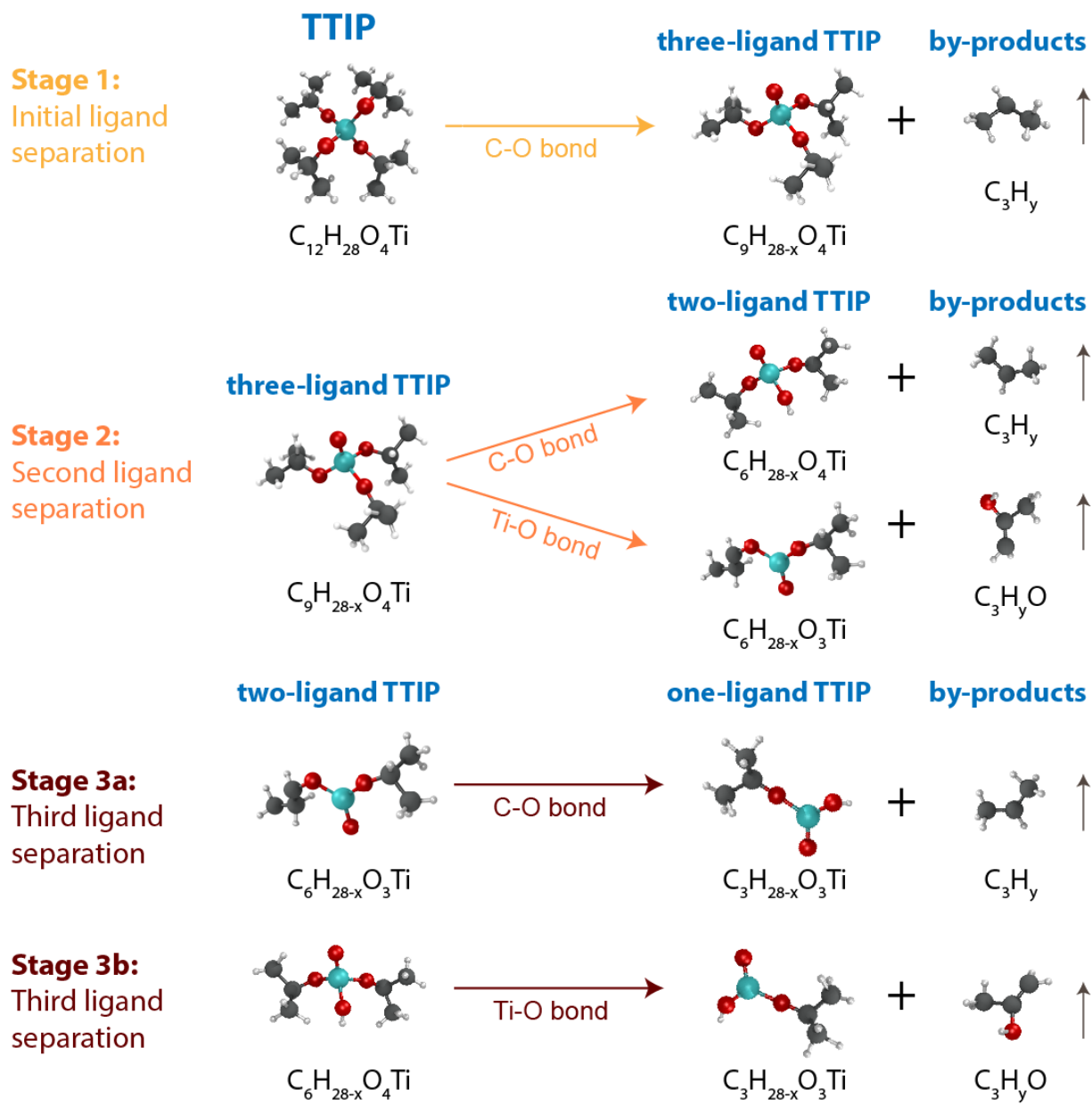

Figure 2. The major events in the decomposition of 25 TTIP molecules show the products and the main by-products formed after each reaction. The left-hand side of each reaction represents the reactants considered in the respective steps; the right-hand side shows the products as well as the volatile by-products formed during the decomposition process. The first step is color coded with yellow, and for the second and third steps, orange and brown were used respectively to distinguish the second and third steps, respectively. This color code was also used for the following figures. Due to the attachment and detachment of H radicals during decomposition, x and y values were not equal.

A total of 25 TTIP molecules ($C_{12}H_{28}O_4Ti$) were randomly placed in three different simulation boxes. It was calculated that there should be 26 gas molecules (in this case, TTIP molecules) in a volume of 100×100×100 Å$^3$ for the substance to be considered as ideal gas based on the ideal gas law. Therefore, our choice of the number of molecules per volume is in good agreement with the condition of an ideal gas at room temperature close to atmospheric pressure. The temperature was linearly ramped from room temperature (initialization step), reached 2000 K at iteration step 1×10$^6$ and was kept at 2000 K throughout the rest of the simulation. The temperature was chosen higher than what is normally used in experiments (~900⁰ C, ~1173 K) to accelerate the thermal decomposition kinetics to balance computational effort and realistically reproduce experimental conditions.

Three different types of chemical reactions were distinguished to categorize all bond dissociation events: 1) C-O bond dissociation 2) Ti-O bond dissociation 3) all other reactions involving C-C and C-H bond fragmentation or formation. For all 75 TTIP molecules simulated, the initial bond dissociation occurred at elevated temperatures during the temperature ramp-up phase of the simulation runs but never below 1800 K. The first organic ligand was liberated from the TTIP molecule by dissociating a C-O bond (category 1), making up close to 90% of the initial ligand separation events. Remarkably, no Ti-O bond dissociation (category 2) was found during the initial step, and in all other 'first bond dissociation' events, 'small fragments' separated off the ligands of TTIP molecules (category 3).

The release of the first ligand was spontaneous and occurred in the absence of any additional reactants. Rather than liberating an alkene group (in the case of TTIP propene ($C_3H_6$)), ReaxFF-MD simulations in the dilute limit revealed the formation of other by-products, specifically propyl radicals ($C_3H_7$), as well. This is in contrast to the *β*-hydride elimination process, by which an alkyl group bonded to a metal center is converted into the corresponding alkene by eliminating a H atom from a *β*-carbon[28], leaving a three-ligand TTIP molecule fragment as a metastable product behind.

Instead, both propene ($C_3H_6$) and propyl ($C_3H_7$) were formed as by-products of the first ligand release stage depicted as $C_3H_x$ in Fig. 2. A three-ligand TTIP molecule ($C_9H_{28-x}O_4Ti$) was formed, in 35% of the cases via a *β*-hydride elimination process (x=6), and 65% of the cases upon propyl (x=7) formation, while the titanium remained tetrahedrally coordinated by oxygen. In this step, H atoms that were observed to be released via *β*-hydride elimination did not reattach to the same molecule but started floating in the simulation boxes and no other hydrogenation during C-O bond dissociations could be observed.

Continuing the thermal decomposition pathways of the three-ligand TTIP after the initial decomposition step, two major routes were identified as second bond dissociation events, see Fig. 2. A C-O bond dissociated releasing $C_3H_x$ (y=6,7) in 60% of the cases, producing a two-ligand TTIP, ($C_6H_{28-x}O_4Ti$ (x=12,13)), either via a *β*-hydride elimination process (62.5%) with the eliminated *β*-H reattaching to the corresponding oxygen or via propyl formation (37.5%) with hydrogenations happening through abstracting H from one of the ligands, while keeping titanium tetrahedrally coordinated by oxygen. In addition, 35% of the second stage of thermal decomposition took part via a Ti-O bond dissociation (category 2). Ti-O bond dissociation was not spontaneous but was observed upon hydrogenation of the three-ligand TTIP via H atoms breaking off of alkyl groups (from the same or neighboring TTIP molecules) and attaching to the remaining oxygens on the molecule. This is yet another route different from the *β*-hydride elimination during decomposition, which required hydrogenation and released $C_3H_yO$ (y=6,8) and produced two-ligand TTIP molecules $C_6H_{28-x}O_3Ti$ (x=13,14,15)). Since oxygen was part of the released ligand, titanium was merely three-coordinated by oxygen. Only 5% of the second bond dissociations happened through sacrificing insignificant fragments during decomposition (category 3).

For the last decomposition step observed in the ReaxFF-MD simulations, two different Ti-containing fragments had to be considered and are distinguished as stages 3a and 3b, respectively. For the three-coordinated two-ligand TTIP (stage 3a), the C-O bonds dissociated in all cases which produced 'one-leigand' molecules ($C_3H_{28-x}O_3Ti$ (x=20,21)) with three-coordinated Ti in addition to $C_3H_y$ (y=6,7) by-products constituting 41% of the third bond dissociation events. While 40% of these events took place via the *β*-hydride elimination and the reattachment of the *β*-H, 60% of the events propyl radicals ($C_3H_7$) were produced instead with hydrogenation. For the tetrahedrally coordinated Ti-containing fragment (stage 3b), the third bond dissociation event occurred between Ti and O producing the same one-ligand TTIP fragment $C_3H_{28-x}O_3Ti$ (x=20,21) and released $C_3H_yO$ (y=6,8) groups. This reaction accounted for the other 41% of the third decomposition step. Again, this Ti-O bond dissociation was triggered by the hydrogenation of the Ti-containing fragment. The remaining 18% of the third bond dissociations occurred as a result of insignificant fragments (category 3) leaving the molecules which increased in portion with increasing temperature during the simulations.

While C-O bond-dissociation events dominated the ligand release initially, the number of the broken Ti-O bonds increased progressively throughout the thermal decomposition process, suggesting that Ti-O bonds require more energy to break when compared with C-O bonds. Ti has the tendency to remain coordinated to oxygen at temperatures lower than 2000 K. A second, important finding is that the first bonds of intact TTIP molecules dissociated spontaneously at this temperature, while the bonds at later decomposition steps required H atoms to bond to oxygen (either by H atoms produced as a part of *β*-hydride elimination during C-O bond dissociation or by the hydrogenation of C-O bond dissociations without *β*-hydride elimination or the hydrogenation of all Ti-O bond dissociations), to initialize further bond dissociation. These H radical atoms made it easier to cleave the bond at every decomposition step except for the first dissociation. Among all the by-products that have been generated at the end of the simulations, the total number of molecules making up more than 1% of the total mole fraction was extracted and is shown below in Table 1. As stated before, 75 TTIP molecules were studied in this section. Based on all by-products formed, it has been concluded that 42 of the C-O bond dissociations have undergone β-hydride elimination producing $C_3H_6$ while 29 have not undergone β-hydride elimination and produced $C_3H_7$. The rest, constituting 18 of them, have been produced as a result of breaking bonds between Ti-O atoms producing $C_3H_yO$ (y=6,8) groups. β-hydride elimination is known to be a reaction type that is favored at lower temperatures[47,48], and $C_3H_6$, which is the only product of β-hydride elimination observed in this study does not explain the decomposition by itself and the β-hydride elimination process is not the only reaction mechanism for thermal decomposition of TTIP in the ReaxFF-MD simulations (Table 1).

| By-product | Number of molecules |
| --- | --- |
| $C_3H_6$ | 42 |
| $C_3H_7$ | 29 |
| $C_3H_8O$ | 11 |
| $C_3H_6O$ | 7 |

Table 1. By-products with mol fractions more than 1% among all other by-products produced as a result of three simulations with 25 TTIP molecules

Starting with 25 TTIP molecules, in the end, the first simulation run contained 77 molecules in the volume, while the second and third runs had 74 molecules and 72 molecules respectively. Generally, $C_3H_y$ and $C_3H_yO$ (y=6,7,8) were the dominant by-products together with three-ligand, two-ligand, and one-ligand TTIP molecules, indicating the significance of C-O and Ti-O bond dissociation processes throughout the thermal decomposition. Although it was possible to assign a map of the types of reactions and estimate their percentages, the simulation box filled up quickly with many different molecules, which were not only products of ligand liberation steps but also resulted from fragmentation of the organic ligands (category 3). Furthermore, it became clear during the analysis of ReaxFF-MD simulations that fragments, in particular hydrogen radicals, altered the thermal decomposition pathways in significant ways when interacting with metastable Ti-containing fragments, which required tracking all individual molecules in the volume considered and made it increasingly difficult to quantify all reaction events individually. Fig. 3 shows the last frame of a simulation where all products and by-products are illustrated. TTIP decomposition took place in the dilute limit by successfully mimicking hMBE thin film growth conditions. There were no $TiO_x$ clusters or any other bond formations that interfered with the decomposition.

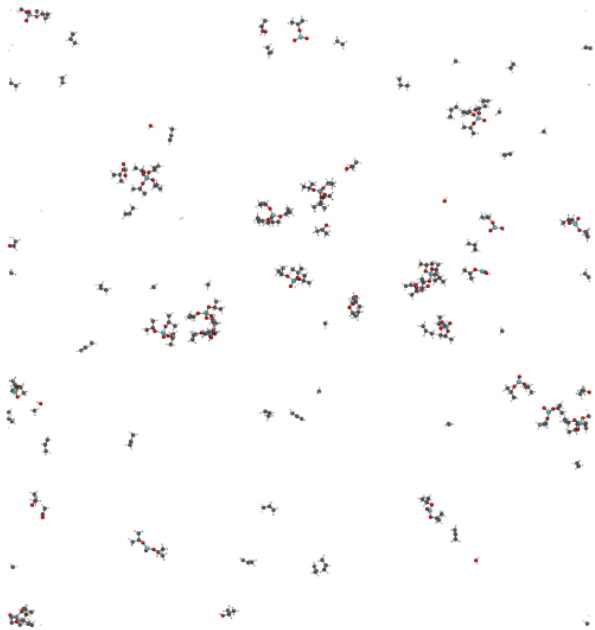

Figure 3. The last frame of a simulation box that initially contained 25 TTIP molecules

Decomposition trends in a single, isolated TTIP molecule

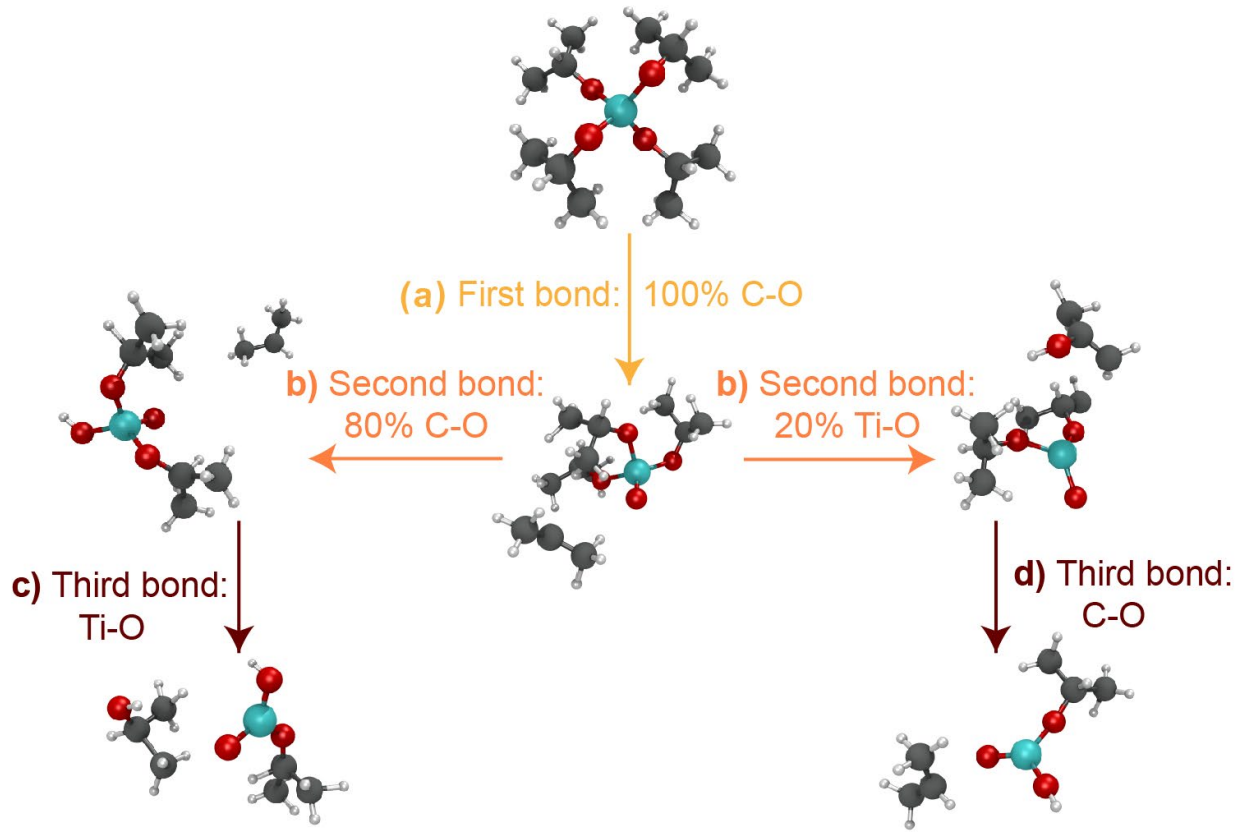

Figure 4. The reaction tree of the decomposition of one TTIP molecule. This overview of bond-dissociation trends in one TTIP molecule shows that (a) the first bond always breaks between C-O (b) 80% of the second bond dissociations occur between C-O and 20% of the second bond dissociations occur between Ti-O atoms. Part (c) represents the case where a third bond dissociation (Ti-O) was observed also (d) indicates another case where a third bond breaks via C-O.

Five different simulations containing a single TTIP molecule in each of them were performed to observe the decomposition trends in a full TTIP ($Ti(OCH(CH_3)_2)_4$) molecule without potential interference from organic ligands or their fragments liberated by adjacent TTIP molecules. The temperature was linearly ramped from room temperature (initialization step), reached 2000 K at iteration step $1\times10^6$ and was kept at 2000 K throughout the rest of the simulation.

The first bond in pristine TTIP molecules broke between C and O atoms in all simulation boxes and did not require hydrogenation (Fig. 4a). Among all the first bond dissociation events, 40% happened through β-hydride elimination of propene and produced $C_3H_6$ with β-H being released to the simulation box while 60% produced $C_3H_7$ molecules without β-hydride elimination and hydrogenation. In the next step, 80% of the bond dissociations of the three-ligand TTIP molecule (second bond, Fig.4b) occurred between C and O in which 50% showed β-hydride elimination with β-H reattaching to the corresponding oxygen while the other half dissociated without β-hydride elimination and produced $C_3H_7$ with hydrogenation. As another route for the second bond dissociation, 20% managed to break the bonds between Ti and O atoms with hydrogenation implying that some amount of oxygen-deficient products is inevitable as temperature increases.

In just two our of the five of the simulation runs, two-ligand molecules were able to further dissociate and liberate a third MO ligand, which is shown in Figs. 4c and 4d. In the former case the Ti-O bond was dissociated forming $TiO_3(C_3H_7)$ with hydrogenation, while in the latter, a C-O bond breaks and the molecule evolves into $TiO_3H(C_3H_7)$ without β-hydride elimination but with hydrogenation. In both cases, the Ti atom is coordinated by 3 oxygen, which suggests that there is an over-oxidized Ti-atom. This overoxidation is enabled by the presence of the last organic ligand, which can potentially become a source of unintentional carbon incorporation. In contrast to the first bond dissociation which did not require hydrogenation, during the second bond dissociation events, C-O bond dissociations occurring without β-hydride elimination and all Ti-O bond dissociations happened in the presence of an additional H radical atom, which gets attached to one of the oxygens. While at most three bond dissociation events were observed in all simulation runs, the total number of steps was not enough to observe the last ligand leaving the TTIP molecule due to the absence of additional H radical atoms or a catalyzing environment to help decompose TTIP further even after increasing the time scale to $3\times10^6$ iteration steps. Furthermore, since the statistics of this section are based on only five simulations, therefore the percentages of observing C-O and Ti-O bond dissociations are slightly different from the previous section with 25 TTIP molecules.

The +4-oxidation state is the most common and stable form of Ti ions, however the oxygen-deficient one-ligand compounds ($TiO_3(C_3H_7)$ and $TiO_3H(C_3H_7)$) that were created at the end prevented the decomposition from moving forward and losing the last attached ligand. Here, the complete thermal decomposition of TTIP requires another partner such as a neighboring TTIP molecule ready to donate H radicals, which are known to accelerate the dissociation, in the proximity of a decomposing TTIP molecule which is a challenging scenario to enable in the dilute limit. Secondly, the complete thermal decomposition can also be accelerated by a substrate providing a relatively electropositive surface to the arriving Ti-containing fragment (in this case $TiO_3(C_3H_7)$ or $TiO_3H(C_3H_7)$), which now becomes attracted by the surface and thus less volatile. Because the surface helps undercoordinated and electronegative oxygen that is prone to establish a link/bond between the Ti-containing fragment and the electropositive surface. This bond formation would then be a chemical rather than a physical bond and would allow for charge transfer, thus helping release the last organic ligand.

Decomposition kinetics of a single TTIP molecule in the presence of additional reactants

From initial ReaxFF simulation runs, it has become clear that hydrogenation reactions are essential at the later stages of the decomposition process. In particular, hydrogen radicals played a significant role in the ligand release, and it is to be expected that the donation and receipt of H radicals in the vicinity of interfaces will play a critical role in the overall decomposition process of the MO molecule. This observation has led us to consider ways to facilitate the decomposition of TTIP. Three different simulation runs were performed to identify to what extent additional fragments such as H radical atoms, $H_2$, and $H_2O$ molecules can be used to accelerate the decomposition kinetics of TTIP. Simulation parameters, such as the iteration number and the applied temperature regime, were kept unaltered to allow a comparison to the first simulation runs.

The temperature was linearly ramped from room temperature (initialization step), reached 2000 K at iteration step $1\times10^6$ and was kept at 2000 K throughout the rest of the simulation. A single TTIP molecule was placed into the simulated volume along with either (i) 5 H radical atoms, (ii) 5 $H_2$ molecules, or (iii) 5 $H_2O$ molecules.

In the cases of $H_2$ and $H_2O$ molecules, the decomposition kinetics of TTIP was unaffected. TTIP molecule started decomposing after $1\times10^6$ steps by itself via a C-O bond and with $\beta$-hydride elimination, in the presence of 5 $H_2$ molecules where $H_2$ molecules stayed the same until the end of the simulation without getting interacted with TTIP. $H_2O$ molecules were similar in this regard and not effective in the process of decomposing a TTIP molecule. Due to the thermodynamically stable nature of $H_2$ and $H_2O$ molecules at the temperatures TTIP decomposition occurs, no reactions were observed between $H_2$ and $H_2O$ molecules and TTIP. 10 H radical atoms, however, increased the decomposition rate by bonding to oxygens at each decomposition step except for the first step which is exactly what has been found in previous simulations.

| Simulation | Iteration number of the first dissociation |
|---|---|
| 1. 25 TTIP | 868,450 |
| 2. 25 TTIP | 944,200 |
| 3. 25 TTIP | 849,500 |
| 4. 1 TTIP + 10 H atoms | 158,500 |
| 5. 1 TTIP | No earlier than 750,000 |

Table 2. Different simulation conditions with changing composition and corresponding iteration numbers of first bond dissociations.

When compared to the cases where TTIP first decomposed at temperatures great than 1800 K and at iteration numbers 800,000 or above (Simulation 1, 2, 3, and 5), a significant speedup in TTIP decomposition can be realized in Simulation 4 with 10 H radical atoms where the TTIP molecule started decomposing ~6.5 times faster at the iteration number 158,500 and roughly at 700 K (Table 2). This indicates the potential of utilizing hydrogen plasmas to accelerate low-temperature TTIP decomposition and oxide growth.

As mentioned before, the time scale that can be achieved in MD simulations is a limiting factor when addressing the questions about TTIP decomposition and implies that there is a need for further investigation. Although MD is a powerful tool for investigating the decomposition trends providing an atomistic picture, it is challenging to compare MD timescales to experiment. A different simulation technique, in which the very small-time steps per iteration (0.25 fs) are not the limiting factor is chosen to address this challenge. Metadynamics[49] simulations have been performed to comprehensively investigate the thermal decomposition pathways and to quantify the energy barriers that are required to overcome during C-O and Ti-O bond dissociations for all Ti-containing MO fragments identified in the previous ReaxFF-MD simulations.

Metadynamics analysis of TTIP

Despite ramping up the temperature to accelerate the reaction kinetics, bond dissociations were fairly rare events during the ReaxFF-MD simulations in the dilute limit. Achievable time scales using typical computational resources were on the order of nanoseconds.[50] Hence, not all possible chemical reactions were observed in a ReaxFF-MD simulation environment. Thus, additional metadynamics calculations were performed to complement the ReaxFF-MD simulations.[51] Metadynamics is a enhanced sampling method used to calculate static properties. In metadynamics, first, collective variables (CVs), the types of bonds in question are identified, and then by dictating a biasing potential as a function of these CVs, the system is assisted in escaping free energy minima by breaking these bonds and visiting new regions in configuration space that would be practically inaccessible in unbiased molecular dynamics simulations.[49,52,53] This repulsive bias discourages the system from revisiting configurations that have been already investigated (i.e. the production of a one-ligand TTIP molecule) which accelerates sampling of the rare events.[54] This method was used in this study to calculate the energy barrier heights of the decomposition steps by 'filling energy wells with computational sand' thereby forcing the system to migrate from one minimum to the next (i.e. from one Ti-containing fragment to another). To calculate the energies of C-O and Ti-O bonds, PLUMED[55,56] was used as a plugin that works with a large number of molecular dynamics codes.

C-O and Ti-O bond energies via metadynamics simulations

Metadynamics simulations were used to estimate the free energies of bond dissociations in a TTIP molecule. The types of bonds in a TTIP molecule were selected as the collective variable (the parameter being sampled over the reaction coordinate), which was visited by the system to calculate the energies needed to break corresponding bonds by applying a biasing potential. Fig. 5 shows the free energies of C-O and Ti-O bonds of TTIP and Ti-containing MO fragments that were discovered to be dissociating in the previous sets of ReaxFF-MD simulations. The first bond was forced to break between one of the four equivalent C-O atoms and the free energy of this event was calculated to be 44 kcal/mol, producing a three-ligand TTIP molecule ($C_9H_{28-x}O_4Ti$) with three ligands left as shown in Fig. 5. Next, the product of the first reaction which was a three-ligand Ti-containing fragment was studied as the reactant in the second step. Since there were two types of bonds, C-O and Ti-O, observed at this stage, both of those bond types were investigated and calculated to have bond energies of 63 kcal/mol and 75 kcal/mol, producing two-ligand $C_6H_{28-x}O_4Ti$ and $C_9H_{28-x}O_3Ti$, respectively.

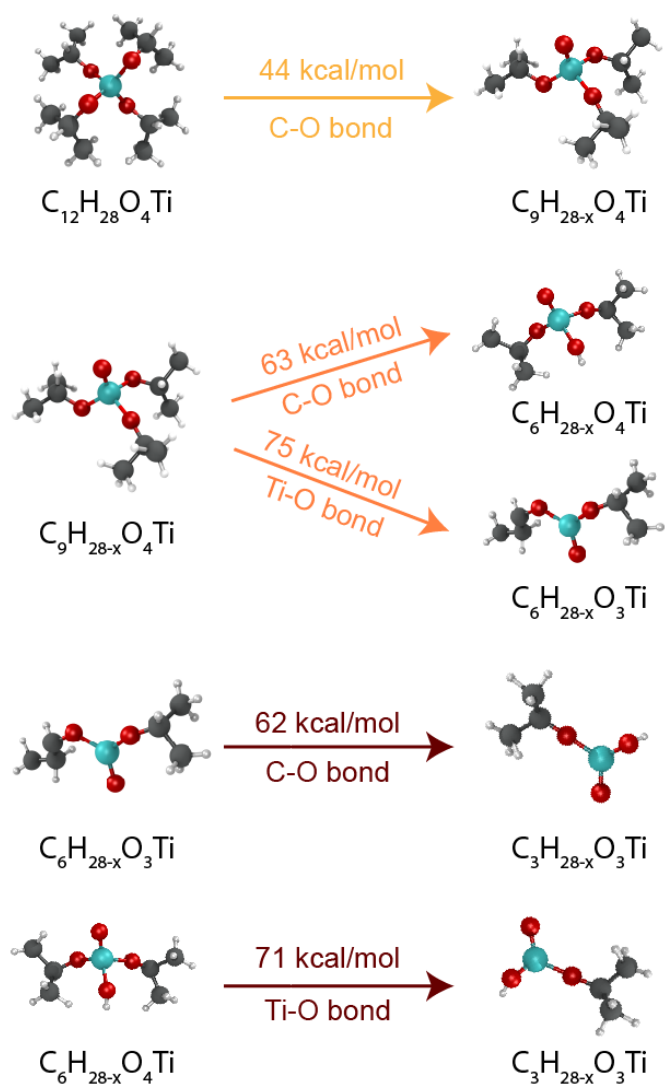

Figure 5. C-O and Ti-O bond energies of each bond that was observed to be dissociating in previously analyzed ReaxFF-MD simulations. Each bond dissociation is illustrated to show the breaking bonds alongside the energies required for each bond dissociation event highlighting that all reactions are endothermic.

The Ti-O bond energy was found to be higher than the C-O bond as predicted in the previous sections. The products of the second step were taken as the reactants of the third stage and the bond energies revealed 62 kcal/mol for the C-O bond and 71 kcal/mol for the Ti-O bond both resulting in $C_3H_{28-x}O_4Ti$. Higher bond energies of Ti-O are in perfect agreement with ReaxFF-MD and further corroborate that Ti-O bond dissociations are harder to achieve. All bond dissociations occurred in the presence of a H radical which bonded to one of the oxygens either due to *β*-hydride elimination or hydrogenation except for the first bond dissociation. Because there was no other molecule than a single TTIP in the simulation box, oxygen atoms scavenged H from the neighboring ligands themselves at each bond dissociation as in the previous simulations except for the first step.

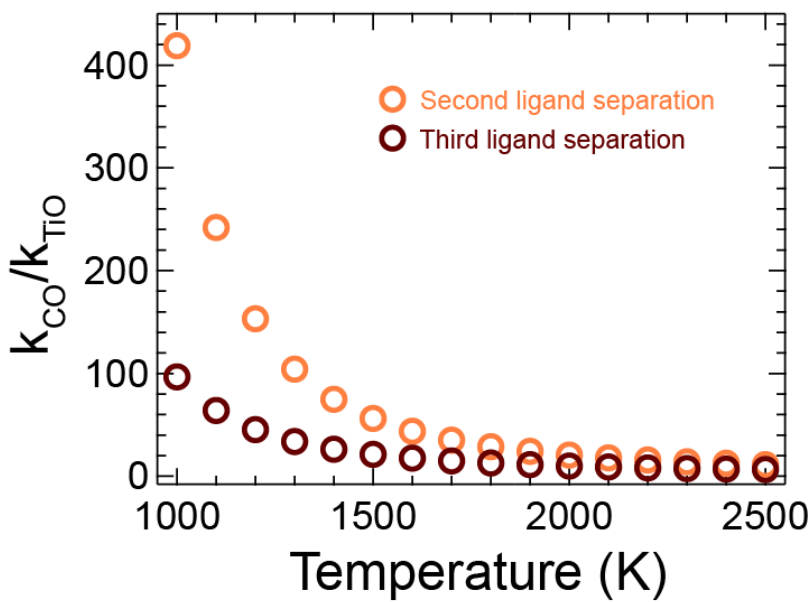

Figure 6. The ratios of the rate constants of C-O and Ti-O bond dissociation reactions ($k_{CO}/k_{TiO}$) based on Arrhenius Law using the bond energies calculated in metadynamics simulations.

By using the free energies calculated via metadynamics and based on the Arrhenius Law, $k_{CO}/k_{TiO}$ was calculated as the ratio of the rate constants of C-O and Ti-O bond dissociations which are the two major events that were detected in all simulation types. At 2000 K, $k_{CO}/k_{TiO}$ for the second ligand separation was calculated to be 20.5 while for the third bond separation this ratio was 9.6 which overall explains the increasing amount of Ti-O bond dissociations at higher temperatures as seen in all previous MD simulations. Since the decomposition of TTIP is a first-order reaction, the free energies of both dissociations were then mapped with respect to temperatures between 1000 K and 2500 K with 100 K increments. As seen in Fig. 6, for both second and third ligand separations, C-O bond dissociations outran the Ti-O bond dissociation at lower temperatures. This finding sheds light on what temperature range should be aimed for controlling bond dissociations. It becomes critical to control such decomposition reactions to increase the amount of C-O bond dissociations instead of Ti-O, in order not to lose oxygen atoms but retain them on the molecule for eliminating oxygen vacancies in the thin films grown with TTIP.

In summary, using a combined approach of ReaxFF-MD simulations and metadynamics calculations, different thermal decomposition pathways of TTIP were investigated. The major conclusions are as follows:

1) In contrast to the conventional assumption, $\beta$-hydride elimination of propene is not the only possible reaction pathway for TTIP decomposition. In addition to C-O bond dissociations, Ti-O bond dissociations can also be observed together with C-O bond dissociations which in general motivates new experiments in the dilute limit to confirm these reaction products.

2) While decomposing, TTIP tends to hold onto the oxygen atoms. The bonds often break between C-O atoms and Ti-O bond dissociations are not observed in the first $1\times10^6$ iteration steps (<2000 K) keeping the oxygens attached to the main product. This clarifies why TTIP can act as an oxygen source to grow titanates without additional oxygen in hMBE systems.

3) A TTIP molecule usually breaks bonds by itself but for further Ti-O bond dissociation, it requires higher temperatures (~2000 K) and scavenges H radicals from the neighboring TTIP molecules.

4) Not all fragments are found to be useful to expedite the decomposition. The presence of H radicals is quite critical to fasten and also facilitate the whole decomposition process while $H_2$ and $H_2O$ have no effect whatsoever.

5) Metadynamics revealed weaker C-O bond energies (44, 59, 63 kcal/mol) compared to Ti-O bonds (75, 78 kcal/mol) irrespective of the metastable state considered, suggesting that the undesired carbon contamination during the MO decomposition process occurs via reabsorption of carbon-containing reactants and incorporation into the material. $k_{CO}/k_{TiO}$ was also calculated based on free energies using the Arrhenius Law and plotted with respect to different temperatures revealing increasing Ti-O bond dissociations at high temperatures. Thus, relatively low temperatures (<1500 K) should be selected for less oxygen-deficient thin films grown by using TTIP.

In conclusion, the overall outcome of this study is enabling new insights to improve MO precursor molecule design with tailored decomposition and reaction pathways to inform the next-generation deposition processes. This becomes very crucial for ultra-scaled electronic devices in which the effect of the substrate's chemical surface state as well as the chemistry of the film growth front needs to be realistically described at the atomic scale instead of relying on simplifying assumptions.

## 3. Methods

Reactive Force Field (ReaxFF) was selected as the force field of the simulations that were performed in this study to investigate the decomposition steps of TTIP molecules using the Amsterdam Density Functional (ADF) package[57] for all simulations. TTIP molecules were simulated in ReaxFF using a previously developed and refined parameter set for Ti, C, O, H.[32–35]

I. <u>ReaxFF MD simulations for TTIP decomposition</u>

a) <u>25 TTIP molecules</u>

Three ReaxFF-MD simulations were performed in the Amsterdam Density Functional (ADF) software where 25 TTIP molecules were generated randomly in each simulation box via ReaxFF. The simulations were performed inside cubic simulation boxes of $100 \times 100 \times 100$ Å$^3$ to ensure a dilute environment for the molecules so that unexpected bond formations can be avoided. Each simulation in this section was performed with a total of $2 \times 10^6$ time steps and a total simulation time of 0.5 ns. Utilizing the NPT (constant number of atoms, pressure, and temperature) ensemble, the pressure was set to be 0 MPa with the Berendsen damping constant of 500 fs. Initial temperature (T) was selected as 300 K with a Berendsen damping constant of 100 fs, and the change in temperature (dT) of 0.0017 K for each time step until $1 \times 10^6$ steps were completed, and 2000 K was reached. A relatively high temperature was selected to accelerate the reaction speed as the time scale (~ns) of MD simulations is limited due to computational costs.[58] For the second half of the simulations, in other words, the remaining $1 \times 10^6$ steps, temperature was kept constant at 2000 K until the end of each simulation.

b) 1 TTIP molecule

To examine the dissociation tendency of TTIP bonds in detail, 5 ReaxFF-MD simulations were performed in the Amsterdam Density Functional (ADF) software where TTIP molecules were generated via ReaxFF. The simulations were performed inside cubic simulation boxes of $100 \times 100 \times 100$ Å$^3$ to ensure a dilute environment for the molecules so that unexpected bond formations can be avoided. Each simulation in this section was performed with a total of $2 \times 10^6$ time steps and a total simulation time of 0.5 ns while utilizing the NPT ensemble where the pressure was 0 MPa with a Berendsen damping constant of 500 fs. Initial temperature (T) was selected as 300 K with a Berendsen damping constant of 100 fs, and the change in temperature (dT) of 0.0017 K for each time step until $1\times10^6$ steps were completed, and 2000 K was reached. A relatively high temperature was selected to accelerate the reaction speed as the time scale (~ns) of MD simulations is limited due to computational costs.[58] For the second half of the simulations, the temperature was kept constant at 2000 K until the end of each simulation.

c) TTIP molecule with additional fragments (H, $H_2$, $H_2O$)

Three ReaxFF-MD simulations were performed in the Amsterdam Density Functional (ADF) software where TTIP molecules were generated together with 10 H radicals, 5 $H_2$, and 5 $H_2O$ via ReaxFF. Each simulation in this section was performed with a total of $2\times10^6$ time steps and a total simulation time of 0.5 ns. An NPT ensemble was used where the pressure was 0 MPa with a Berendsen damping constant of 500 fs. Initial temperature (T) was selected as 300 K with a Berensend damping constant of 100 fs, and the change in temperature (dT) of 0.0017 K for each time step until $1\times10^6$ steps were completed, and 2000 K was reached. A relatively high temperature was selected to accelerate the reaction speed as the time scale (~ns) of MD simulations is limited due to computational costs.[58] For the second half of the simulations, in other words, the remaining $1\times10^6$ steps, temperature was kept constant at 2000 K until the end of each simulation. The ReaxFF-MD simulations were performed inside cubic simulation boxes of $100\times100\times100$ Å$^3$ to ensure a dilute environment for the molecules so that unexpected bond formations can be avoided.

II. Metadynamics

a) C-O bond energy calculation:

To calculate the free energies of each C-O bond dissociation in a TTIP molecule (Fig. 7b), PLUMED was used as a plugin that works with a large number of molecular dynamics codes. A Berendsen thermostat was selected at 2000 K and the damping constant was set at 100 fs inside simulation boxes with the dimensions of 100×100×100 Å³ in which a full, three-ligand, two-ligand TTIP molecules were placed randomly. The bonds that were selected to be broken were indicated in the input script based on their numbers as shown in Fig. 7b. The simulation was performed for $10\times10^6$ steps with a time step of 0.25 fs and the sampling frequency of 200 with the ReaxFF force field implemented in the AMS suit.[57]

b) Ti-O bond energy calculation:

To calculate the free energies of each Ti-O bond dissociation in a TTIP molecule (Fig. 7a), PLUMED was used as a plugin that works with a large number of molecular dynamics codes. A Berendsen thermostat was selected at T=2000 K and a damping constant of 100 fs in simulation boxes of 100×100×100 Å³ in which a full, three-ligand, two-ligand TTIP molecules were placed randomly. The bonds to be broken were indicated in the input script based on the atom numbers shown in Fig. 7a. The simulation was performed for $10\times10^6$ steps with a time step of 0.25 fs and a sampling frequency of 200 in the ADF software based on ReaxAMS.

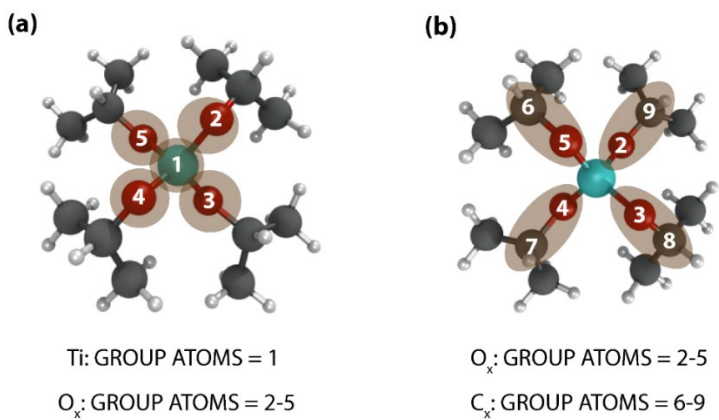

Figure 7. Ti-O (a) and C-O (b) bonds are enumerated in the Plumed input script to calculate the free energies of each bond

This part of the work was carried out using the open-source, community-developed PLUMED library[55], version 2.4[56].

**Acknowledgment**

Computations for this research were performed on the Pennsylvania State University's Institute for Computational and Data Sciences' Roar supercomputer. B. F. Y. and R. E.-H. wish to acknowledge National Science Foundation through DMR-1905861 for data acquisition. The authors would also like to thank the US Department of Energy, Office of Science, Basic Energy Sciences, through Award Number DE-SC0020145 as part of the Computational Materials Sciences Program.

**Data availability**

The data that support the findings of this study are available from the corresponding author, B. F. Y., upon reasonable request.

**Competing Interests**

The authors declare no conflict of interest.